\documentstyle[epsfig,aps,twocolumn]{revtex}
\newcommand{\rv}{{\bf r}}

\newcommand{\beq}{\begin{equation}}
\newcommand{\eeq}{\end{equation}}
\newcommand{\bea}{\begin{eqnarray}}
\newcommand{\eea}{\end{eqnarray}}

\renewcommand{\>}{\rangle}

\renewcommand{\[}{\left[}
\renewcommand{\]}{\right]}
\newcommand{\commentout}[1]{{}}

\begin{document}
\draft
\title{Creating vortex rings and three-dimensional skyrmions in Bose-Einstein
condensates}
\author{J. Ruostekoski$^{1}$ and J. R. Anglin$^2$}
\address{$^1$Department of Physical Sciences, University of Hertfordshire,
Hatfield, Herts, AL10 9AB, UK\\
$^2$Institute for Theoretical Atomic and Molecular Physics,\\
Harvard-Smithsonian Center for Astrophysics, Cambridge MA 02135\\
}
\date{\today}
\maketitle

\begin{abstract}
We propose a method of generating a vortex ring in a Bose-Einstein
condensate by means of electromagnetically-induced atomic transitions. Some
remnant population of atoms in a second internal state remains within the
toroidal trap formed by the mean field repulsion of the vortex ring. This
population can be removed, or it can be made to flow around the torus (i.e.
within the vortex ring). If this flow has unit topological winding number,
the entire structure formed by the two condensates is an example of a
three-dimensional skyrmion texture. \
\end{abstract}

\pacs{03.75.Fi,05.30.Jp}

\preprint{}

\begin{figure}
\begin{minipage}{3.9cm}
\epsfig{width=3.9truecm,
file=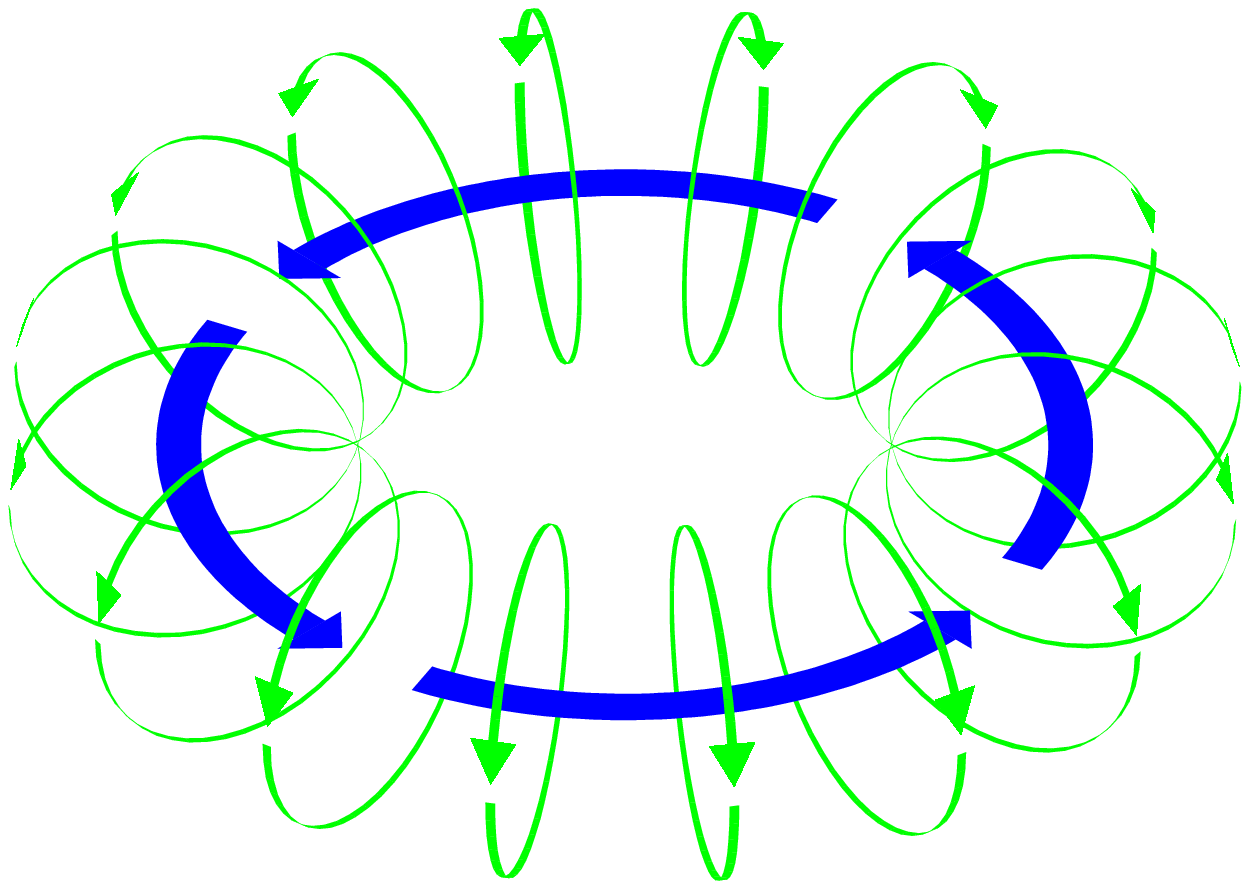}
\end{minipage}
\begin{minipage}{4.3cm}
\epsfig{width=4.3truecm,file=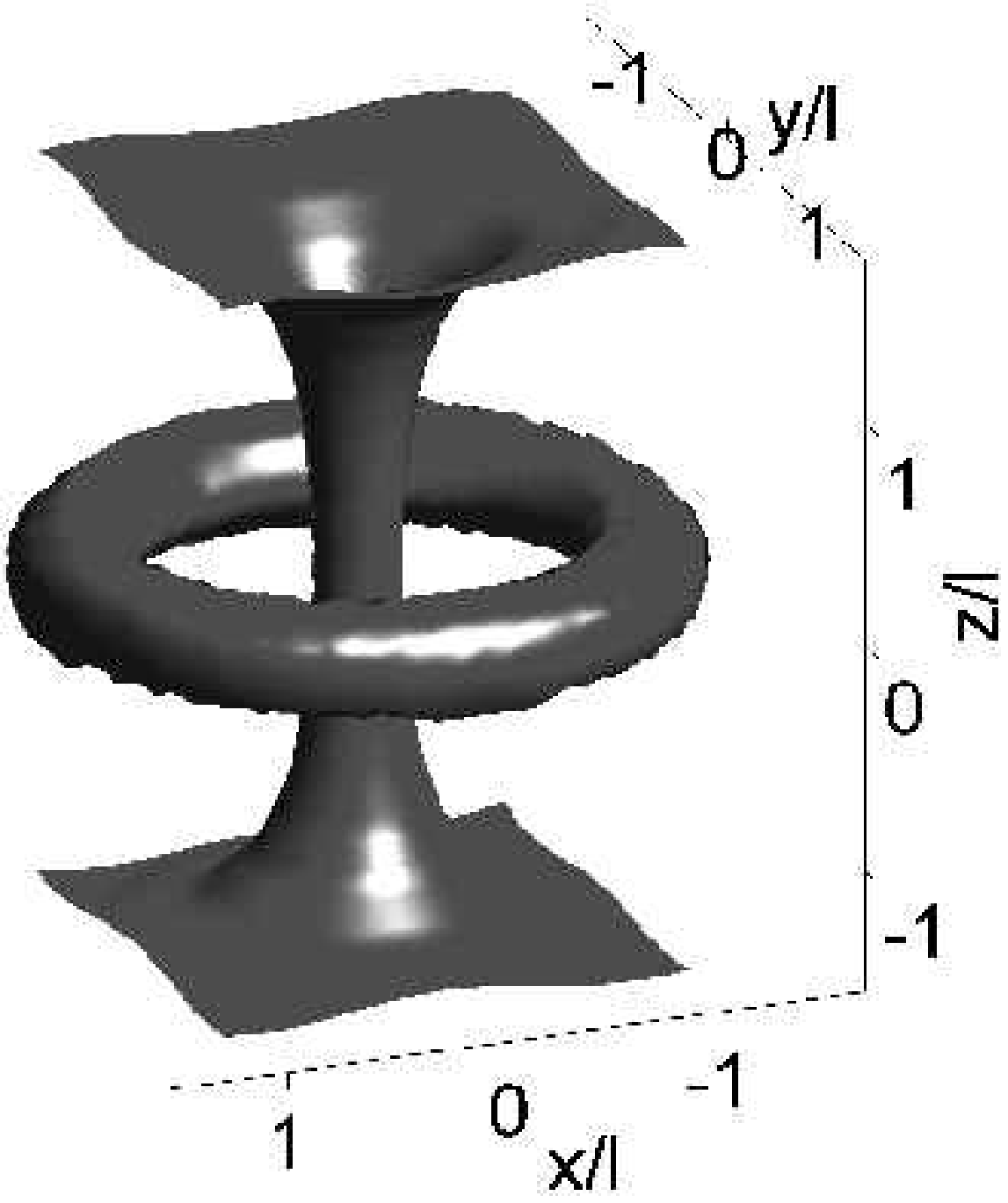}
\end{minipage}
\caption{Schematic illustration of a 3D skyrmion (left), as a vortex ring
containing a superflow, and a constant surface plotting of the 3D skyrmion
(right) obtained from the numerical simulations. We show the core
region of the vortex line and the ring by
displaying the atom densities for $ |\psi_1|^2= |\psi_2|^2 =1.2
\times 10^{-4}/l^3 $.
 } \label{f1}
\end{figure}

With the recent controlled creation of vortices \cite{MAT99,MAD00} in
trapped atomic Bose-Einstein condensates (BECs), and evidence for vortex
nucleation above a critical flow velocity \cite{RAM99}, the dynamics of
topological structures in weakly interacting superfluids has become an
active subject of experimental research. In this paper we propose methods of
preparing a vortex ring within a BEC, allowing the controlled study of a
vortex structure away from surface effects \cite{AND00}.
The vortex ring is created by
means of driving electromagnetic (em) fields inducing transitions between
internal atomic levels. The em field couples to the relative phases between
the levels, in such a way as to transfer angular momentum to the atoms. With
an appropriately chosen superposition of em fields the local axes of this
effective rotation form a closed circular loop, producing a vortex ring in
the BEC.  We also consider the stability of the vortex ring, and
identify a way to use a second atomic component of the BEC to increase the
stability of the
ring against collapsing under its own string tension. Conversely, the
vortex ring acts as a trap for the second component, providing purely
atom-optical confinement of cold atoms. Finally, the combined two-component
structure may be identified as a three-dimensional (3D) {\it skyrmion}
\cite{SKY61}, shown in Fig.~\ref{f1}. This nonsingular texture is of interest
in its own right, as a topological structure beyond the simple vortex.

Topological defects such as vortices and monopoles possess cores at which
the order parameter is singular, and indeed such defects are characterized
by winding numbers defined on paths or surfaces that enclose the singular
cores. The simplest topological defect relevant to superfluid physics
is that of a current flowing around a closed 1D path parametrized
by the angular coordinate $\phi$. The simple mapping
$\alpha(\phi)=\phi$, from physical space into order parameter
space, where $\phi$ is defined on a 1D circle, has `winding number' one,
which will be unchanged if $\alpha(\phi)$ is continuously deformed.
Topologically nontrivial configurations are also possible without
singular cores: {\it Textures} are defined by the way in which the
compact order parameter space is `stretched over' physical space.  An SU(2)
order parameter, for instance, takes values on a sphere. \ One can
`puncture' this sphere, stretch the resulting hole to infinity, and so
`spread' the order parameter space over a 2D plane.  Such a mapping, or any
continuous deformation of it, is known as a 2D (or `baby') {\it skyrmion}. \
Interpreting the order parameter sphere as the spin orientation of a spinor
condensate, one can realize a 2D skyrmion as a coreless point vortex (or in
3D, a coreless vortex line) \cite{Ho,MAR00}, such as has been created
experimentally at JILA \cite{MAT99}.

The full, 3D skyrmion results from taking an $S^{3}$ order parameter space
and stretching it over $R^{3}$ in a similar way \cite{Stoof}. An $S^{3}$
order parameter space is afforded most simply by a two-component BEC,
whose interactions effectively fix the total density $|\psi _{1}|^{2}+|\psi
_{2}|^{2}=\rho $ of the two complex macroscopic wave functions $%
\psi _{1},\psi _{2}$. The low energy degrees of freedom therefore reside on
a 3-sphere of angles $0<\alpha,\beta \leq \pi$, $0<\gamma \leq
2\pi $ (a 3-sphere is a 3D subset in 4D, such that $\sum_{n=1}^4 x_n^2= R^2$).
For spherical polar coordinates in physical space, a 3D skyrmion is
given by any continuous deformation of the mapping $\gamma (r,\theta ,\phi
)=\phi ,\beta (r,\theta ,\phi )=\theta ,\alpha (r,\theta ,\phi )=\lambda (r)$
for monotonic $\lambda $ with $\lambda (0)=0,\lambda (\infty )=\pi $. The
topologically invariant winding number
\begin{equation}
W=\frac{1}{2\pi ^{2}}\int \!d^{3}x\,\sin ^{2}\alpha \,\sin \beta \,\det
\left( \frac{\partial \alpha ^{i}}{\partial x^{j}}\right) \;,
\end{equation}
which obviously vanishes for the ground state of the two-species BEC,
equals one for the skyrmion. \ (The topological invariance of $W$ is not
hard to understand: because the determinant in the integrand is precisely
the Jacobian for the change of variables from $x^{i}=(x,y,z)$ to $\alpha
^{i}=(\alpha ,\beta ,\gamma )$, the integral is simply the volume of the
unit 3-sphere, regardless of the precise form of $\alpha ^{i}(x^{j})$.)

We can obtain a physical picture of the 3D skyrmion by writing out the
macroscopic wave functions:
\begin{eqnarray}  \label{skyrmsol}
\left(\matrix{\psi_1(\rv)\cr \psi_2(\rv)}\right) &\equiv& \ \sqrt{\rho(r)}%
\left(\matrix{-i\sin\alpha\sin\beta\exp(i\gamma)\cr
\cos\alpha-i\sin\alpha\cos\beta}\right)  \nonumber \\
&=& \sqrt{\rho(r)}\left(\matrix{-i\sin[\lambda(r)]\sin\theta\exp(i\phi)\cr
\cos[\lambda(r)]-i\sin[\lambda(r)]\cos\theta}\right)\,.
\end{eqnarray}
With this realization of the skyrmion, $|\psi_1|^2$ vanishes on the $z$
axis and at infinity, and is concentrated in a toroidal region. On the other
hand, $|\psi_2|^2$ vanishes on the circle $\theta={\frac{\pi}{2}},r=
\lambda^{-1}({\frac{\pi}{2}})$. Expanding $\psi_2$ about any point on this
circle, we find $\psi_2 \propto \delta r + i\delta\theta$, indicating that
the nodal circle of $\psi_2$ is a vortex line. Hence component $\psi_2$
forms a vortex ring, within whose core $\psi_1$ resides and flows
azimuthally with winding number one. Of course, any continuous deformations
of Eq.~(\ref{skyrmsol}) will still be a skyrmion, and the two-fluid
interpretation of the skyrmion will always provide a useful intuitive
picture. Moreover, unlike in the case of wave function monopoles~\cite{wfm},
changing the atomic basis that defines the two fluids does not lead to a
different picture of the skyrmion, but simply produces the same combination
of vortex ring and current, with a different spatial orientation.

We propose a method of creating a 3D skyrmion in a trapped BEC using em
fields, in a scheme motivated by viewing the skyrmion as this combination of
vortex structures. Since several good methods have been proposed for
generating simple vortices, and some of these have already been successfully
implemented, engineering the state of $\psi _{1}$ presents no
new difficulties of principle. Creating the vortex ring state in $\psi _{2}$
is more of a challenge. Although vortex rings may be formed through
instabilities, as of a dark soliton in an insufficiently narrow trap \cite
{AND00}, a more controlled method of preparing them may also be desirable,
in order to study their rich range of motion and excitation.

We consider em-induced coherent scattering of atoms between different
internal levels. By superposing different driving em fields we can construct
a Rabi amplitude with a spatially-dependent phase profile. In particular,
nodes of the em field amplitude may be topological singularities of the em
phase. The coupling of the em field to the phase of the coherent matter wave
allows the topological singularities of the em fields to be {\it imprinted}
on the matter field. We demonstrate the preparation process of topological
objects by numerically studying this coherent matter-wave dynamics.

The dynamics of the BECs with the Rabi coupling between the levels $|i\>$
and $|j\>$ follows from the coupled Gross-Pitaevskii equation (GPE)
\begin{equation}
i \hbar \dot\psi_i = {\large ( H_i^0+\delta_i +\sum_k \kappa_{ik} |\psi_k|^2
) \psi_i+\hbar\Omega_{ij}^* \psi_j\,.}  \label{gpe}
\end{equation}
Here the kinetic energy and the trapping potential are introduced in $H_i^0$%
:
\begin{equation}
H_i^0\equiv -{\frac{\hbar^2}{2m}}{\bbox \nabla}^2+{\frac{1}{2}} m\omega_i^2
(x^2+\alpha_i^2 y^2+\beta_i^2 z^2)\,.
\end{equation}
We have also defined the interaction coefficients $\kappa_{ij}\equiv
4\pi\hbar^2 a_{ij}N/m$. Here $a_{ii}$ denotes the intraspecies scattering
length in internal level $|i\>$ and $a_{ij}$ ($i\neq j$) stands for the
interspecies scattering length. The Rabi frequency, $\Omega({\bf r})$,
describes the strength of the coupling between the two internal levels. The
total number of BEC atoms, the atomic mass, and the detuning of the em
fields from the resonance are denoted by $N$, $m$, and $\delta_j$,
respectively.

\begin{figure}
\vbox{  \vspace{-10mm}\hbox{ \epsfig{width=3.7truecm,file=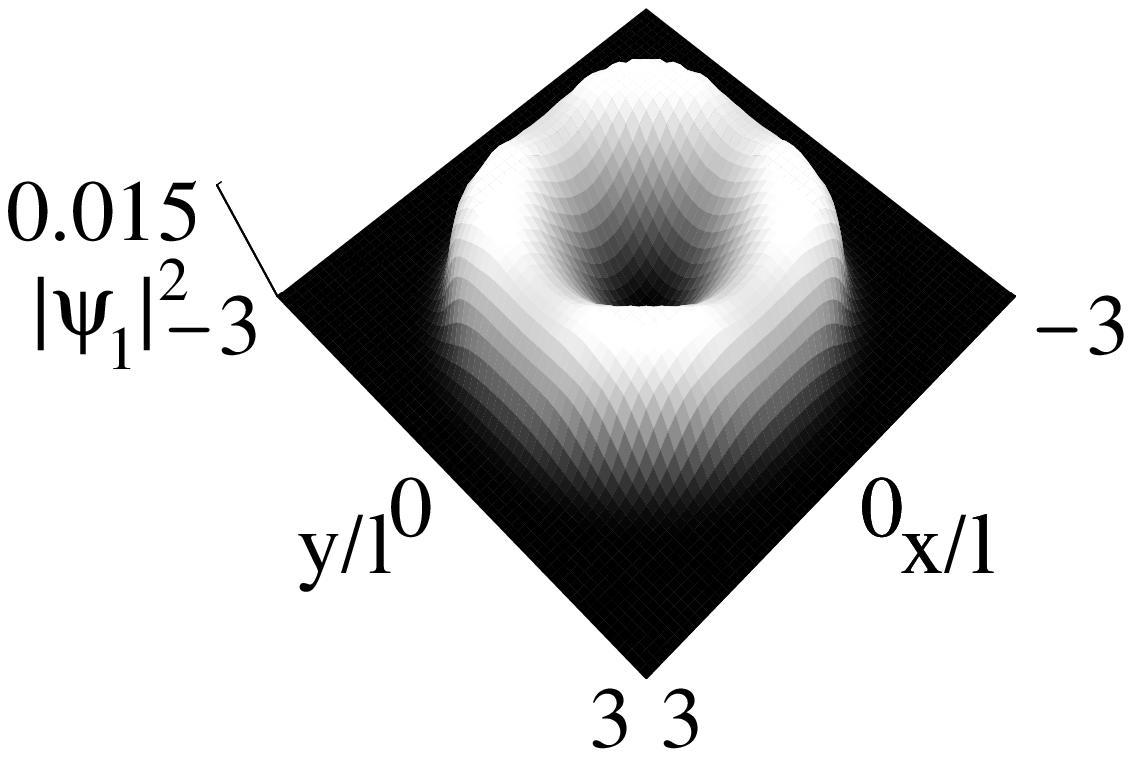}
\epsfig{width=3.7truecm,file=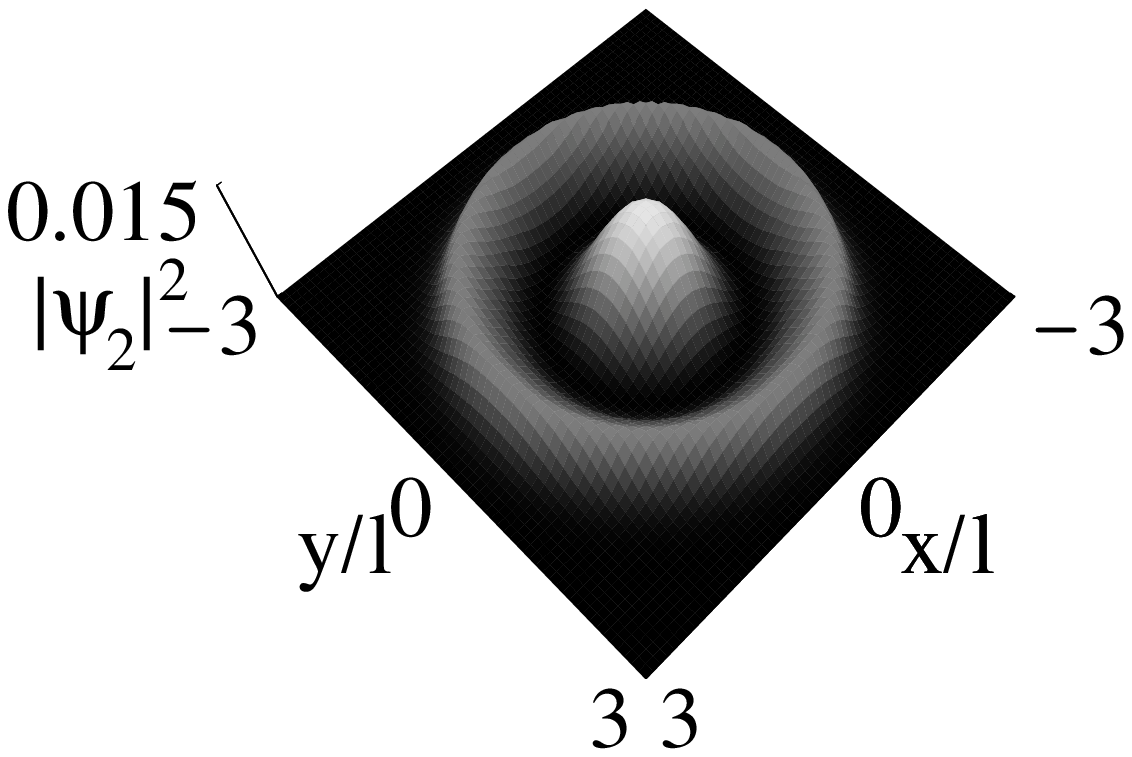}} \vspace{-7mm}\hbox{
\epsfig{width=3.7truecm,file=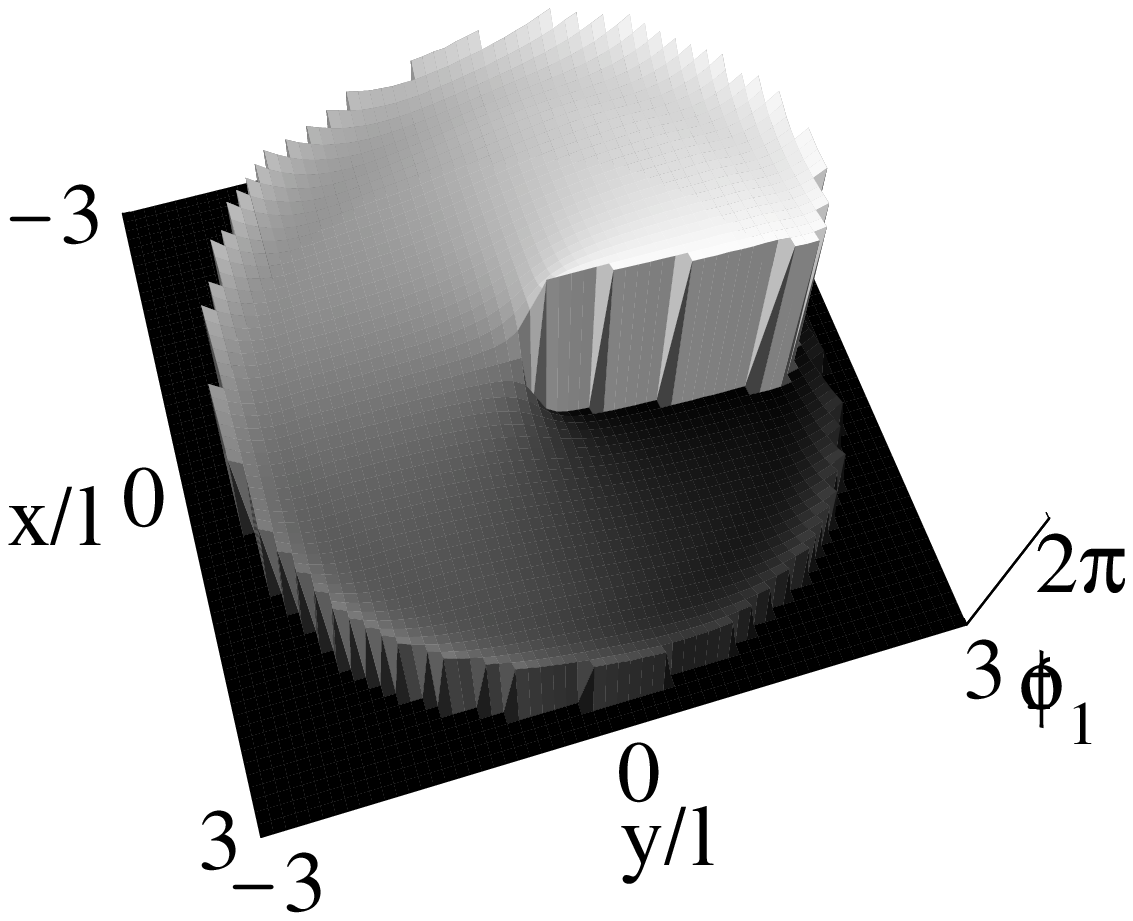}
\epsfig{width=3.7truecm,file=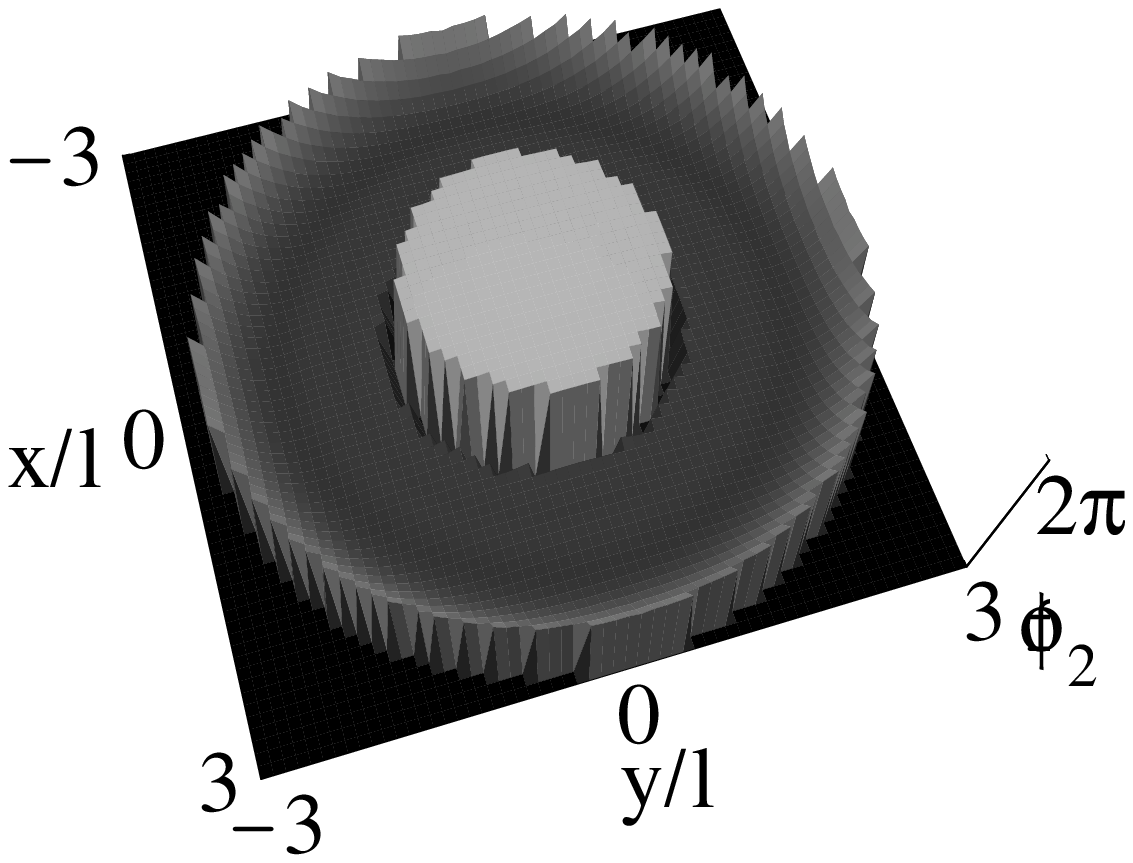}}} \caption{A 3D skyrmion
in the $xy$ plane. We show the density $|{\psi}(x,y)|^2$ and the
phase $\phi={\rm arg}[ {\psi}(x,y)]$ profiles of the wave function
in the $xy$ plane. Here ${\psi}(x,y)\equiv\int dz \psi(x,y,z)$.
The density profile (upper left) and phase profile (lower left) of
atoms in level $|1\>$ display a single vortex line with $2\pi$
phase winding around the core, and vanishing density at the center
of the core. The vortex line occupies a toroidal region where the
density of the vortex ring of atoms in level $|2\>$ (upper right)
vanishes. The phase distribution of the vortex ring (lower right)
in the $xy$ plane is flat.} \label{f2}
\end{figure}

Several papers \cite{BOL98,MAR97,DUM98,WIL99,DOB99,RUO00b,MAR00} have
proposed methods of injecting vorticity into BECs by means of the
interaction between em and matter fields. In our adaptation of these methods
to create a 3D skyrmion, the simplest part is the step of producing the
`inside' component $\psi_1$ with a quantized circulation about the $z$ axis.
This can be achieved by transferring population from an initial internal
state $|0\rangle$ using the Rabi frequency $\Omega_{01}({\bf r})$,
\begin{equation}
\Omega_{01}({\bf r}) = \Omega_0 [ \sin(k_1x)-i\sin(k_1 y)%
 ]\cos(k_2z) \,,  \label{ome}
\end{equation}
with a similar set of parameters as in Ref.~\cite{RUO00b}. In particular, $%
k_1^{-1}$ and $k_2^{-1}$ are both larger than the size of the sample. The
purpose of the $\cos(k_2z)$ is to confine $\psi_1$ along the $z$ direction.
Note that in this scheme the Rabi frequency could represent a two-photon
transition, with one photon from two orthogonal standing waves in the $xy$
plane and the other from a standing wave along the $z$ direction.
Alternatively, the $\cos(k_2z)$ could simply represent collimation of the $xy
$ beams, which would drive a single photon transition by themselves. The
topological singularity in the em field at $x=y=0$ results in a single
quantum of circulation about this axis, in the component $\psi_1$, as
displayed in Fig.~\ref{f2}. One may also generate additional terms in Eq.~(%
\ref{ome}). For instance, the standing wave along the $z$ axis could be
further collimated in the $xy$ plane, limiting the spatial extent of $\psi_1$.

We propose two different methods of creating the more complicated part of
the skyrmion, the closed vortex ring in $\psi_2$. The first is an adaptation
of the method of Ref.~\cite{MAT99}, and makes use of two internal levels and
the circulating state $\psi_1$ once it has been generated. To add the vortex
ring requires a Rabi field coupling from state $|1\rangle$ to state $%
|2\rangle$, which is a phase-coherent superposition of a standing em field
and a Gaussian beam, both along the $z$ axis:
\begin{equation}
\Omega_{12}({\bf r})={\Omega_0}
\[
1-\eta^2-\exp{[-\rho^2/\xi^2]}+i A\sin(k_3z)
\]
\,,  \label{ome2}
\end{equation}
with $A=2\eta/(k_3\xi)$ and $\rho\equiv(x^2+y^2)^{1/2}$. We are interested
in $k_3|z|\ll1$, $\rho\ll \xi$, and $\eta\ll 1$ in such a way that $%
|k_3(\rho-\eta\xi)|\ll 1$. Then we obtain
\begin{eqnarray}
\Omega_{12}({\bf r})/\Omega_0 &\simeq& \rho^2/\xi^2-\eta^2+iAk_3z  \nonumber
\\
&\simeq& 2\eta/\xi(\rho-\eta\xi)+iAk_3z  \nonumber \\
&=& Ak_3[(\rho-\eta\xi)+iz] \,.
\end{eqnarray}
The phase of $\Omega_{12}({\bf r})$ has a form of the quantized circulation
with unit winding number around the closed ring $\rho=\eta\xi$ and $z=0$.
Alternatively, the Gaussian beam resulting in the exponential term in Eq.~(%
\ref{ome2}) may also be replaced by two standing em fields: $%
[\cos(2x/\xi)+\cos(2y/\xi)]/2$.

For this first method, the field of Eq.~(\ref{ome2}) is tuned away from
resonance. A localized shifting field is then applied as in Ref.\cite{MAT99}%
, to `turn on' the transition in a small region, which is then swept around
the trap to create a full vortex ring, and also undo the azimuthal phase
winding of the $|1\rangle$ atoms, so that the population shifted into $%
|2\rangle$ will have no angular momentum about the $z$ axis. While this
scheme takes advantage of an already demonstrated technology, it has the
disadvantage that it tends to produce a highly excited skyrmion state, which
may (as discussed below) then evolve self-destructively.

We therefore propose, in our second method, to couple atoms with em fields
from level $|0\rangle$ to level $|2\rangle$, at the same time as we generate
the simple vortex state in level $|1\rangle$. We need a Rabi field coupling
from $|0\rangle$ to $|2\rangle$, of the same spatial form as $\Omega_{12}$
of Eq.~(\ref{ome2}), as well as the $\Omega_{01}$ field of Eq.~(\ref{ome}).
This scheme requires control of an additional internal state $|0\rangle$,
but with a weak coupling it allows adiabatic transfer of the atomic
populations \cite{DUM98}, producing a ground state skyrmion.

For computational simplicity we demonstrate the three-level preparation
process by considering two short nonadiabatic Rabi pulses: The first Rabi
pulse of the form (\ref{ome2}) generates the vortex ring in level $|2\rangle$
from the initial ground state population in level $|0\rangle$, and then the
second pulse with the configuration (\ref{ome}) generates the trapped
current in level $|1\rangle$ from the remaining atoms in level $|0\>$.
Our numerical
approach uses three states and can still generate an excited skyrmion, but
it is amenable to fully 3D simulation, and our simulation indicates that the
method can work. A careful study of an adiabatic population transfer will
require a realistic model for the dissipation of the BECs, as well
significant computational resources, and must be left for future
investigation.

In our numerical simulations we assume that the traps are isotropic, $%
\alpha_i=\beta_i=1$, and that the trapping frequences are equal for all
components. We choose the wave number of the em fields $k_1=k_2=k_3=2\pi/(8l)
$, with $l \equiv [\hbar/(m\omega)]^{1/2}$, $\delta_i=0$, $\eta=0.04$%
, $\xi=30l$, and $\kappa_{ij}=100\hbar\omega l^3$. The number of atoms is
determined by the previous relations as $N=25l/(\pi a)$.

\begin{figure}
\vbox{ \vspace{-10mm}\hbox{ \epsfig{width=3.7truecm,file=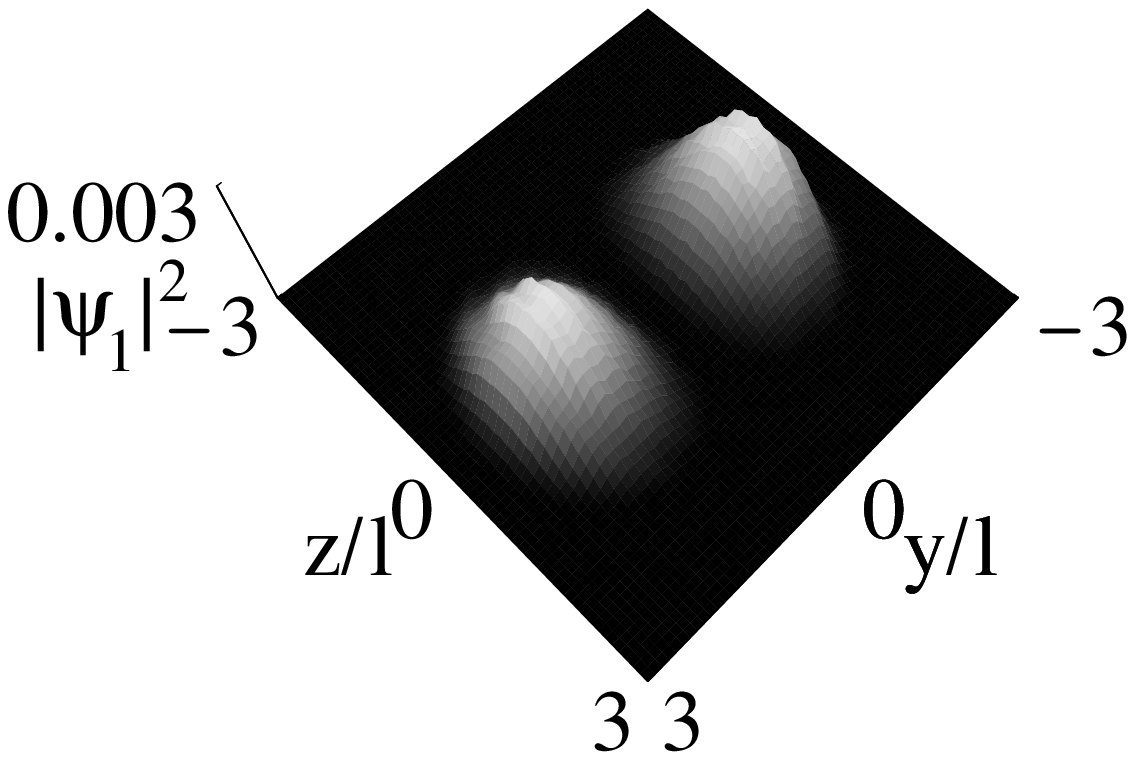}
\epsfig{width=3.7truecm,file=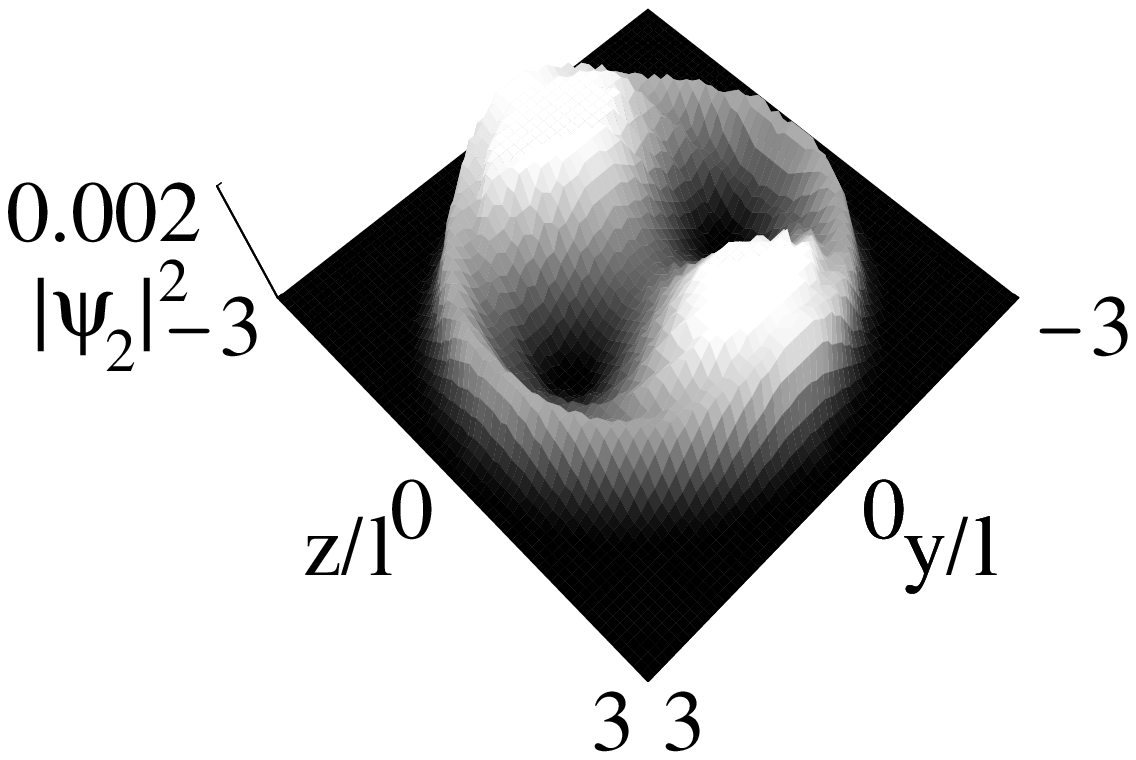}} \vspace{-7mm}\hbox{
\epsfig{width=3.7truecm,file=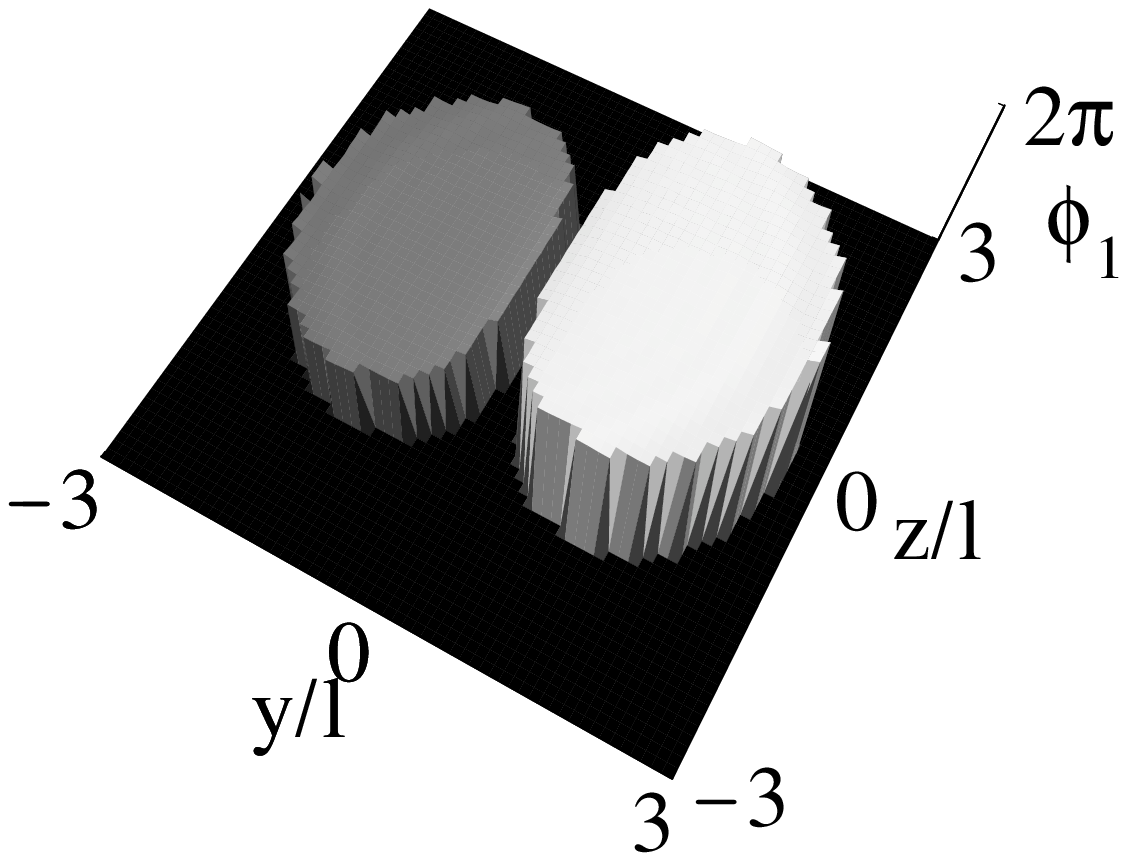}
\epsfig{width=3.7truecm,file=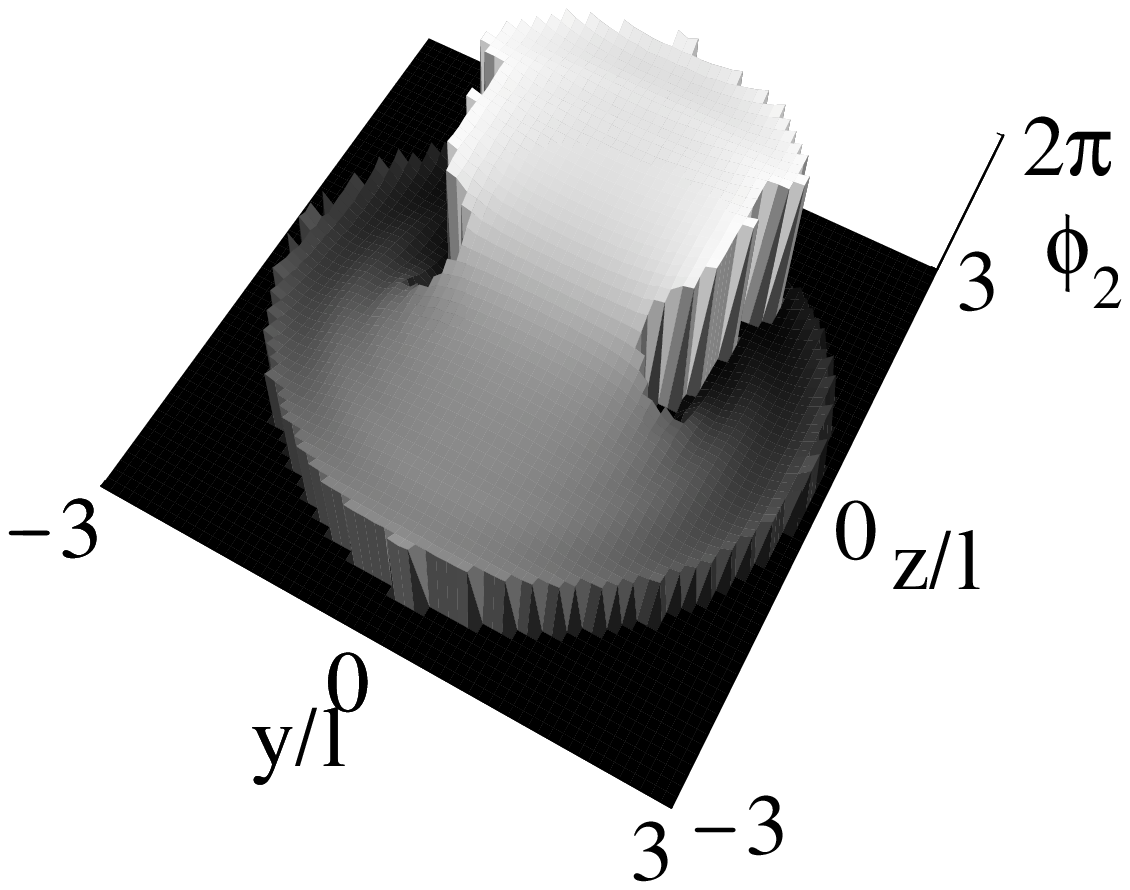}}} \caption{A 3D skyrmion
in the $yz$ plane. We display the density $|\psi(x=0,y,z)|^2$ and
the phase $\phi={\rm arg}[\psi(x=0,y,z)]$ profiles of the wave
function. The density of the single vortex line in level $|1\>$
(upper left) occupies the toroidal region surrounded by the vortex
ring in level $|2\>$ (upper right). The phase profile of the
vortex ring (lower right) displays the quantized rotation around
the ring. } \label{f3}
\end{figure}

We integrate the GPE on a spatial grid of $128^{3}$ points, using a
split-operator method. The stability of the nonlinear evolution is improved
by simultaneously propagating multiple copies of the wave function.
Relatively small number of time steps (less than one thousand) provide
a sufficient accuracy making the integration executable. Even with up-to-date
workstations the optimized simulations can take several hours.
Figures~\ref{f1}-\ref{f3} represent the 3D skyrmion in the case where we
have first created the vortex ring in level $|2 \>$ by applying the
Rabi frequency $\Omega _{12}({\bf r})$ [Eq.~(\ref{ome2})], with
$\Omega _{0}=5000\omega $ for $0< t\omega\leq 0.04$.
This is followed by creating a vortex line in level $|1\>$
by the driving $\Omega _{01}({\bf r})$ [Eq.~(\ref{ome})],
with $\Omega_0 =80\omega $ for $0.04< t\omega \leq 0.05$.
With the chosen set of parameters a large population remains
in $|0\>$. This remnant population may be coupled
out of the trap immediately after the em pulses as in the recent vortex
experiments \cite{MAT99}.
Due to the shortness of the pulses the interactions of the
atoms in level $|0 \>$ do not have time to affect the structure of
the skyrmion. On the other hand, for an efficient adiabatic population
transfer the nonlinearity can be a problem. However, the
techniques to overcome this were developed in Ref.~\cite{DUM98}.

The effort of creating such an intricate structure as a
vortex ring, let alone a 3D skyrmion, is obviously only justified if its
structure can then be observed. As with the filled vortex lines of JILA, the
interior BEC in state $|1\rangle $ will enable detection of the density and
phase profile of the skyrmion.
An additional advantage of the skyrmion, as opposed to an empty vortex
ring or to a vortex ring filled with a nonrotating interior BEC, is that it
is less unstable against shrinking to zero radius. The kinetic and
interaction energy of a vortex line gives it an effective string tension
tending to shorten the vortex line,
and in the case of a vortex ring, the ring's radius $R$ (not
its core thickness!) is energetically unstable to shrinking. \ In an
infinite, homogeneous background, and in the absence of dissipation, the
first-order dynamics of a vortex line means that a vortex ring will not
actually shrink, but will instead move along its axis, at a velocity
determined by its radius. But if its radius approaches its core thickness, a
vortex ring can be annihilated (it is not topologically stable), and an
excited vortex ring may also possibly twist or pinch, and break up into
smaller rings. Moreover, motion at a constant speed cannot continue in a
finite, inhomogeneous BEC. Dissipation is also present in
experimental conditions, and although straight vortices have been found to
have a long lifetime before floating out of the BEC cloud, the energy
slope due to string tension may be much steeper than that due to vortex
buoyancy (on the healing length scale rather than trap scale). \ Filling the
vortex core lowers the string tension, and hence extends the lifetime of the
filled vortex structure in comparison with that of an empty ring.

Although in other field theories skyrmions are often unstable against
shrinking to zero size without the topological change of defect nucleation,
the separate conservation of both atomic species in our case forbids
this decay channel. This does not prevent the vortex ring from collapsing,
but it does mean that the component $|1\>$ filling the ring cannot be
eliminated.  Furthermore, since the filling has one unit of
angular momentum, there is a centrifugal barrier inhibiting
the shrinking. As it tends to pin the vortex ring down, the ring
may only collapse via the tunneling of atoms through $\psi_1$.
Although the vortex rings are dynamically stable \cite{JAC00},
numerical evolution in imaginary time shows that filled vortex rings, and
even skyrmions, are still energetically unstable, on long enough (imaginary)
time scales.  Skyrmions are topologically stable as long as the order
parameter is well defined everywhere; but they can be destroyed by the
nucleation of a defect. \ Our simulations indicate that, while the energy
decreases slowly but monotonically, the vortex ring in $\psi _{2}$ can pull
through the state $|1\rangle $ if its inner radius approaches the
healing length, so that a healing-length radius circle of zero total density
nucleates. However, the decay via the tunneling of atoms
through $\psi_1$ becomes increasingly slow with the larger
occupation in $|1\>$ \cite{com1}. A vortex ring with a nonrotating filling can,
if its radius becomes too small, annihilate itself without significantly
lowering the total density at any point.

Without dissipation, numerical evolution in real time shows that skyrmions
as well as empty vortex rings do move through the background cloud, but
instead of simply exiting the cloud as they approach its surface, they
generate a background flow in the rest of the cloud which is able to draw
them back towards the center. Much more extensive numerical studies
will be needed to determine the stability of the resulting oscillation.
Further analytical and numerical studies of skyrmion and vortex ring
dynamics are clearly warranted, but in this paper we have shown that
engineering and study of these advanced BEC structures should be
feasible with current techniques.

This research was supported by the NSF and by EPSRC. We thank
C.S.~Adams, J.~Dziarmaga, H.T.C.~Stoof, and W.H.~Zurek for
valuable discussions, and gratefully acknowledge the hospitality of ITAMP
at the Harvard-Smithsonian Center for Astrophysics.


\begin{references}
\bibitem{com1} A very large BEC in level $|1\>$ filling the vortex ring may
also annihilate the vortex ring by pushing it through the outer boundary
of the atomic cloud as a result of the mean-field repulsion.
\end{references}
\end{document}